\documentclass[12pt]{iopart}

\usepackage{graphicx}

\begin{document}

\title[]{Laser interferometry with translucent and absorbing mechanical oscillators}

\author{D Friedrich$^{1}$, H Kaufer$^1$, T Westphal$^1$, K Yamamoto$^{1,\footnotemark[3]}$, A Sawadsky$^1$, F Ya Khalili$^2$, S L Danilishin$^2$, S Go\ss ler$^1$, K Danzmann$^1$ and R Schnabel$^{1}$}

\address{$^1$ Max-Planck-Institut f\"ur Gravitationsphysik (Albert-Einstein-Institut) and Institut f\"ur Gravitationsphysik, Leibniz Universit\"at Hannover}

%\address{$^2$ Institute for Cosmic Ray Research, The University of Tokyo, 5-1-5 Kashiwa-no-Ha, Kashiwa, Chiba 277-8582, Japan}

\address{$^2$ Department of Physics, Moscow State University, Moscow RU-119992, Russia}

\ead{Daniel.Friedrich@aei.mpg.de}

\begin{abstract}
The sensitivity of laser interferometers can be pushed into regimes that enable the direct observation of quantum behaviour of mechanical oscillators. In the past, membranes with subwavelength thickness (thin films) have been proposed as high-mechanical-quality, low-thermal-noise oscillators. Thin films from a homogenous material, however, generally show considerable light transmission accompanied by heating due to light absorption, which typically reduces the mechanical quality and limits quantum opto-mechanical experiments in particular at low temperatures. In this work, we experimentally analyze a Michelson-Sagnac interferometer including a translucent silicon nitride (SiN) membrane with subwavelength thickness. We find that such an interferometer provides an operational point being optimally suited for quantum opto-mechanical experiments with translucent oscillators. In case of a balanced beam splitter of the interferometer, the membrane can be placed at a node of the electro-magnetic field, which simultaneously provides lowest absorption and optimum laser noise rejection at the signal port. We compare the optical and mechanical model of our interferometer with experimental data and confirm that the SiN membrane can be coupled to a laser power of the order of one Watt at $1064\,$nm without significantly degrading the membrane's quality factor of the order 10$^6$, at room temperature.
% memT=69.63%
% s-pol 51.4/48.6
% p-pol 24.06/75.94
% faraday 92.49
% 0.0373; -0.0011
\end{abstract}
\footnotetext[3]{Present address: Institute for Cosmic Ray Research, The University of Tokyo, 5-1-5 Kashiwa-no-Ha, Kashiwa, Chiba 277-8582, Japan}

\maketitle

\section{Introduction}
Quantum fluctuations of a light field couple to the motion of macroscopic test mass mirrors via momentum transfer of reflected photons, thus leading to back action noise for a position measurement \cite{Braginsky78, Thorne78, Braginsky, Chen11}. Forthcoming 2nd generation interferometric gravitational wave detectors such as Advanced LIGO \cite{AdvLIGO}, Advanced Virgo \cite{AdvVirgo}, GEO-HF \cite{GeoHF} and LCGT \cite{LCGT} will be limited by quantum noise in most of their detection band. While the detector's signal to shot noise ratio at high Fourier frequencies can be improved with higher laser power, quantum radiation pressure noise will become a limiting noise source at low frequencies \cite{Caves80}. However, the observation of quantum radiation pressure will enable to test the principle of back-action noise in a continuous position measurement. 

The strength of opto-mechanical coupling increases with the amount of light power that is used to sense the test mass position and its susceptibility to the radiation pressure force, in particular with smaller masses. Therefore, today's fabrication techniques for micro-mechanical oscillators have opened new possibilities to study the coupling of light and mechanical devices as reviewed in Ref.~\cite{Kippenberg08}. 

One major adversary for high precision experiments targeting the quantum regime is thermal noise caused by mechanical dissipation related to optical multilayer coatings \cite{Levin98,Harry02}. However, avoiding such coatings comes at the expense of low reflectivity, which raises the need of novel interferometer topologies. Therefore, thin silicon nitride (SiN) membranes placed in a high finesse cavity \cite{Thompson08, Jayich08, Sankey10} are being investigated in order to observe quantization of an oscillators mechanical energy via non-linear opto-mechanical coupling. Besides the low effective mass of about $100\,$ng, SiN-membranes provide high mechanical quality factors exceeding $10^6$ at $300\,$K and $10^7$ at $300\,$mK \cite{Zwickl08}.       

If a semi-transparent component is used (e.g. SiN membranes with a reflectivity $\leq 40\,\%$ at a laser wavelength of $1064\,$nm) as common end mirror for the two arms of a Michelson interferometer the transmitted light forms a Sagnac interferometer. This Michelson-Sagnac interferometer is compatible with advanced interferometer techniques such as power-recycling and signal-recycling \cite{Meers88,Strain91}, which potentially increase opto-mechanical coupling as investigated in Ref.~\cite{Yamamoto10}. It was shown that quantum radiation pressure noise can dominate thermal noise by a factor of three around its eigenfrequency for an incident power of $1\,$kW, which in principle can be realized with power-recycling, or for an incident power of $1\,$W if signal-recycling is adopted. Since in either case the membrane needs to be cooled to a temperature of $1\,$K, optical absorption in the membrane has to be minimized. 

Here, we demonstrate that the lowest optical absorption can be achieved for mechanical oscillators with subwavelength optical thickness placed in the node of a standing wave that is inherent to the Michelson-Sagnac interferometer. Experimentally we identified optical absorption by changes in the eigenfrequency of a SiN membrane, which was accompanied by a decrease of its mechanical quality factor. We further show that placing the membrane in a node is compatible with operating the interferometer at its dark fringe, which will enable the implementation of advanced interferometer techniques.

\section{Light field amplitudes in a Michelson-Sagnac interferometer}
\begin{figure}[ht]
\begin{center}
\includegraphics[width=12cm]{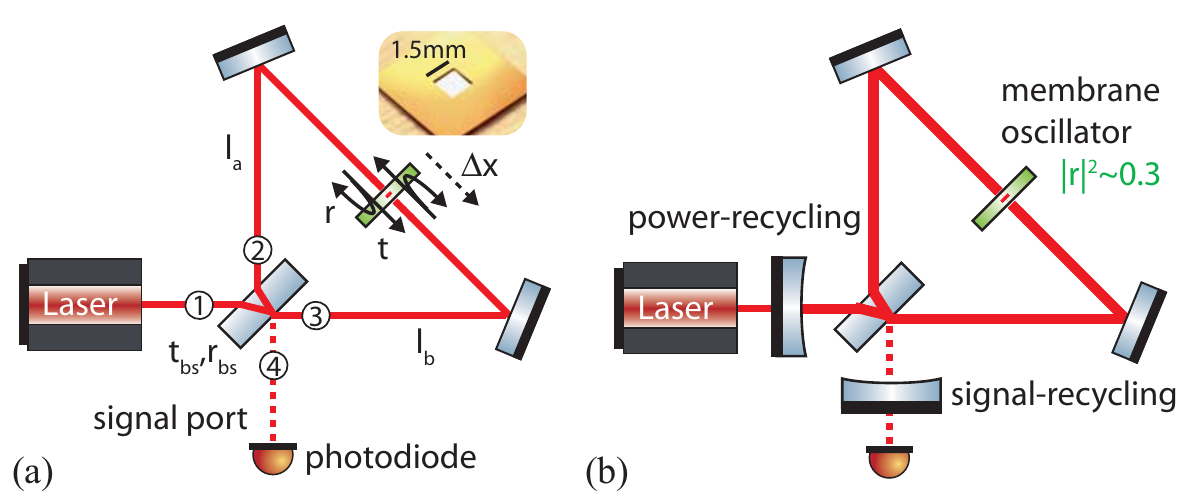}
\caption{(a) Sketch of a Michelson-Sagnac interferometer with a translucent membrane as common end mirror for the two arms of a Michelson interferometer. A displacement $\Delta x$ of the membrane leads to differential arm length change of $\Delta l=l_\mathrm{a}-l_\mathrm{b}=2 \Delta x$ that can be sensed at the signal port. The transmitted light forms a Sagnac interferometer, which has a dark signal port for a 50/50 beam splitter independent of the mirror position. (b) The combined Michelson-Sagnac interferometer operated on its dark fringe enables the use of advanced interferometer techniques such as power-recycling and signal-recycling to increase opto-mechanical coupling as theoretically investigated in \cite{Yamamoto10}.} \label{fig:MiSaIfo}
\end{center}
\end{figure}

A displacement $\Delta x$ of the common end mirror in a Michelson-Sagnac interferometer (Michelson-Sagnac ifo) (see Fig.~\ref{fig:MiSaIfo}(a)) causes a differential arm length change of $\Delta l=l_\mathrm{a}-l_\mathrm{b}=2\Delta x$ thus leading to light power variations at the output ports. With respect to advanced interferometer techniques such as power- and signal-recycling \cite{Meers88,Strain91} as indicated in Fig.~\ref{fig:MiSaIfo}(b) operating the interferometer at its dark fringe is essential. 

In this section we show that the tuning of the Michelson interferometer (Michelson ifo) via a displacement of the membrane is related to the position of the membrane in a standing-wave, originating from the counterpropagating, transmitted light fields that form a Sagnac interferometer (Sagnac ifo). Therefore, the light fields (amplitude and phase) to the interferometers signal port will be derived first.

\begin{figure}[ht]
\begin{center}
\includegraphics[width=12cm]{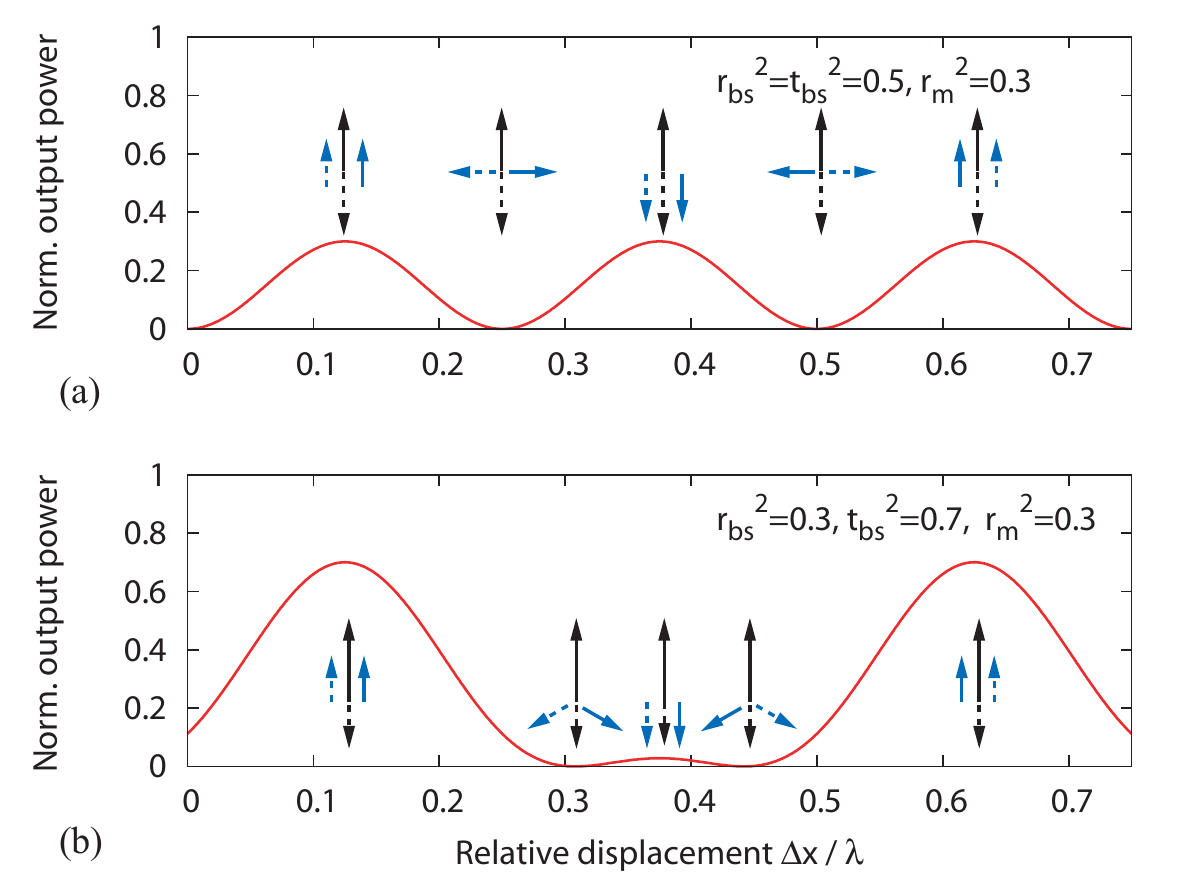}
\caption{Interference of the interferometer's Michelson (blue arrows) and Sagnac (black arrows) light field amplitudes at the signal port. The resulting output power $|a_\mathrm{out}|^2$ (see Eq.~(\ref{eq:MiSaamp})) is shown as a red line, normalized to the input power, for different splitting ratios of the central beam splitter. The phasors illustrate the interference of the combined interferometer. While the light fields transmitted through the membrane (black arrows) are constantly out of phase by $180\,$deg, the reflected ones (blue arrows) are counterrotating for an end mirror displacement $\Delta x$. In case of not too extreme splitting ratios the amplitudes of the Michelson interferometer can compensate a residual amplitude from the Sagnac interferometer.} \label{fig:fringes}
\end{center}
\end{figure}

For the phase relations of reflected and transmitted light fields at the central beam splitter we follow Ref.~\cite{Zeilinger81}. With the numbering of ports given in Fig.~\ref{fig:MiSaIfo}(a) the complex amplitude reflectivity and transmissivity of a lossless beam splitter $r_{ij}=r_\mathrm{bs} \mathrm{exp}(i\Theta_{ij})$ and $t_{ij}=t_\mathrm{bs} \mathrm{exp}(i\Theta_{ij})$, respectively, are restricted by $r_{12}=r^*_{34}$ and $t_{13}=-t^*_{24}$ to ensure energy conservation. Assuming the beam splitter further to be reciprocal ($\Theta_{ij}=\Theta_{ji}$)  these relations are fulfilled by the following phase relations
\begin{eqnarray}
\Theta_{12}=-\Theta_{34} \quad \mathrm{and} \quad \Theta_{13}=-\Theta_{24}\pm \pi,
\end{eqnarray}
which enable to set phases for reflection and transmission independent of each other.
The complex coefficients of the common end mirror are written in the same manner as $r=r_\mathrm{m} \mathrm{exp}(i\Theta_\mathrm{rm})$ and $t=t_\mathrm{m} \mathrm{exp}(i\Theta_\mathrm{tm})$ and will be investigated further in the next section for a translucent mirror with non-zero absorption. The accumulated phases for propagating light fields of wavelength $\lambda$ in a Michelson-Sagnac ifo with arm lengths  $l_\mathrm{a}$ and $l_\mathrm{b}$ leaving at the signal port are at a glance
\begin{eqnarray}
\Theta_\mathrm{sa1}= \Theta_\mathrm{12}+ \Theta_\mathrm{34}+\Theta_\mathrm{tm} +k(l_\mathrm{a}+l_\mathrm{b}),\\
\Theta_\mathrm{sa2}= \Theta_\mathrm{13}+ \Theta_\mathrm{24} +\Theta_\mathrm{tm} +k(l_a+l_\mathrm{b}),\\
\Theta_\mathrm{mi1}= \Theta_\mathrm{12}+ \Theta_\mathrm{24} +\Theta_\mathrm{rm} +2kl_\mathrm{a},\\
\Theta_\mathrm{mi2}= \Theta_\mathrm{13}+ \Theta_\mathrm{34} +\Theta_\mathrm{rm} +2kl_\mathrm{b},
\end{eqnarray}
with the wavenumber $k=2\pi/\lambda$. Consequently the sum of all light field amplitudes at the signal port is given by
\begin{eqnarray}
a_\mathrm{out}=a_\mathrm{in} \left( t_\mathrm{m} r_\mathrm{bs}^2 e^{i \Theta_\mathrm{sa1}} + t_\mathrm{m} t_\mathrm{bs}^2 e^{i\Theta_\mathrm{sa2}} + r_\mathrm{bs} t_\mathrm{bs} r_\mathrm{m}e^{i \Theta_\mathrm{mi1}} + r_\mathrm{bs} t_\mathrm{bs} r_\mathrm{m}e^{i \Theta_\mathrm{mi2}}\right)\\
= \underbrace{a_\mathrm{in} t_\mathrm{m} e^{i \Theta_\mathrm{sa1}}  \left( r_\mathrm{bs}^2- t_\mathrm{bs}^2\right)}_\mathrm{Sagnac\;interferometer} +  \underbrace{a_\mathrm{in} r_\mathrm{bs} t_\mathrm{bs} r_\mathrm{m} e^{\frac{i}{2} (\Theta_\mathrm{mi1}+\Theta_\mathrm{mi2})} 2\cos\left(\frac{\Theta_\mathrm{mi1}-\Theta_\mathrm{mi2}}{2}\right) }_\mathrm{Michelson\;interferometer}. \label{eq:MiSaamp}  
\end{eqnarray}
The differential phase of the Sagnac ifo is $\Theta_\mathrm{sa1}-\Theta_\mathrm{sa2}=\pm \pi$. Hence, in case of a 50/50 beam splitter ($r_\mathrm{bs}^2=t_\mathrm{bs}^2=0.5$) these light fields cancel and the resulting normalized output power reads 
\begin{eqnarray}
P_{50/50}/P_\mathrm{in}= |a_\mathrm{out}/a_\mathrm{in}|^2= r_\mathrm{m}^2 \cos^2\left(k(l_\mathrm{a}-l_\mathrm{b})+\Theta_{12}-\Theta_{13}\pm \pi/2 \right),  
\end{eqnarray}
which is minimal (dark fringe condition) for differential arm lengths $\Delta l =l_a-l_b$ of  
\begin{equation}
k \Delta l=(\Theta_{13}-\Theta_{12} \pm m\pi) \quad \mathrm{with}\quad m=0,1,2,\ldots \quad. \label{eq:darkfringe}
\end{equation}
In case of non 50/50 splitting ratios the residual but constant amplitude of the Sagnac ifo needs to be considered. These basic results are illustrated in Fig.~\ref{fig:fringes} based on Eq.~(\ref{eq:MiSaamp}) for two different splitting ratios and a power reflectivity of the membrane $r_\mathrm{m}^2 =0.3$. 
For not too extreme splitting ratios the amplitudes of a detuned Michelson ifo (counter-rotating blue phasors) are able to cancel the residual amplitude of the Sagnac ifo (not rotating black phasors).

\subsection{Standing wave in a Michelson-Sagnac interferometer}
The two counterpropagating light fields in the interferometer, transmitted through the membrane, form a standing wave in terms of electrical field strength. An anti-node (maximal electric field amplitude) will be present when both fields have accumulated equal phases (at the interferometer's center) or in integer ($m=0, 1, 2,\ldots$) distances of half a wavelength from that point which can be written as
\begin{eqnarray}
\Theta_{12}+k(l_\mathrm{a} \pm \frac{m}{2}\lambda)=\Theta_{13}+k(l_\mathrm{b} \mp \frac{m}{2}\lambda)\\
\Rightarrow k(l_\mathrm{a}-l_\mathrm{b})=\Theta_{13}-\Theta_{12} \mp 2m \pi,\label{eq:antinode}
\end{eqnarray}
while nodes (minimal electric field amplitude) are described by
\begin{eqnarray}
\Theta_{12}+k(l_\mathrm{a} \pm \frac{2m+1}{4}\lambda)=\Theta_{13}+k(l_\mathrm{b} \mp \frac{2m+1}{4}\lambda)\\
\Rightarrow k(l_\mathrm{a}-l_\mathrm{b})=\Theta_{13}-\Theta_{12} \mp (2m+1) \pi.\label{eq:node}
\end{eqnarray}
Comparing these relations with Eq.~(\ref{eq:darkfringe}) one finds that the latter two correspond to the dark fringe condition for the signal port in case of a 50/50 beam splitter. We would like to note that this result is independent of the actual beam splitter phases, since they have not been specified so far. In case of other splitting ratios a superposition of a standing and a travelling wave (partial standing wave) will be present in the interometer, but with minima and maxima in electrical field amplitude at the same mirror positions. Hence, only for 'roughly balanced' beam splitting ratios nodes and anti-nodes will coincide with a dark signal port. In the next section the implications of the standing wave to optical absorption of a translucent mirror will be quantified.

\section{Optical absorption of a translucent mirror}
\begin{figure}[ht]
\begin{center}
\includegraphics[width=12cm]{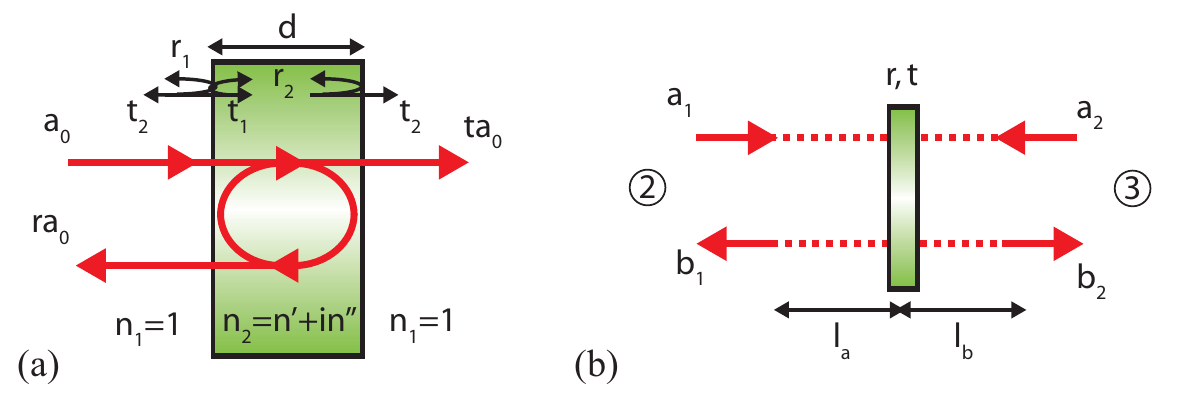}
\caption{(a) Multiple interference model used to calculate the complex reflection and transmission coefficients ($r$, $t$) of a translucent material with thickness $d$ and non-zero absorption Im($n_2) >0$ surrounded by vacuum $n_1=1$. (b) Membrane position in the Michelson-Sagnac interferometer with respect to the counterpropagating light fields $a_1$ and $a_2$.} \label{fig:membrane}
\end{center}
\end{figure}

The optical properties of a translucent material with complex index of refraction $n_2=n^{'}+i n{''}$ (see Fig.~\ref{fig:membrane}(a)) 
surrounded by a non-absorbing material Im($n_1)=0$ can be derived by a multiple interference model with the well known Fresnel equations for reflection and transmission under normal incidence at the material boundaries \cite{Born}
\begin{eqnarray}
r_1 = \frac{n_1-n_2}{n_1+n_2}, \quad r_{2}=\frac{n_2-n_1}{n_1+n_2}, \quad t_1 = \frac{2n_1}{n_1+n_2}, \quad \textrm{and} \quad t_2 = \frac{2n_2}{n_1+n_2}. 
\end{eqnarray}
The incident light field $a_{0}$ is partly reflected $r_{1}$ and transmitted $t_{1}$ at the surface. The transmitted light field undergoes several internal reflections $r_{2}$. After each transit of optical thickness $n_{2}d$ a part is coupled out via transmission $t_{2}$. Summing up all fields one arrives at the membrane's complex amplitude reflectivity and transmissivity  
\begin{eqnarray}
r_ & = & r_{1}+t_{1}t_{2}r_{2}e^{2ikn_2 d}\sum^{\infty}_{m=0}{\left( r_{2}^2e^{2ikn_2 d}\right)^m}=   \frac{r_{1}+r_{2} e^{2ikn_2 d}}{1-r_{2}^2 e^{2ikn_2 d}}, \\
t & = & t_{1}t_{2}e^{ikn_2 d}\sum^{\infty}_{m=0}{\left( r_{2}^2e^{2ikn_2 d}\right)^m} = \frac{t_{1}t_{2} e^{ikn_2 d}}{1-r_{2}^2 e^{2ikn_2 d}}, \label{eq:airy}
\end{eqnarray}
which simplifies for a layer surrounded by vacuum ($n_1=1$) to 
\begin{eqnarray}
r & = & -\frac{(n_2^2-1)\sin(kn_2 d)}{2in_2 \cos(kn_2 d)+(n_2^2+1)\sin(kn_2 d)},\\
t & = & \frac{2in_2}{2in_2\cos(kn_2 d)+(n_2^2+1)\sin(kn_2 d)}.
\end{eqnarray}
For a translucent mirror in the Michelson-Sagnac ifo one has to consider two counterpropagating incident fields ($a_{1}$, $a_{2}$) as depicted in Fig.~\ref{fig:membrane}(b). If we generalize for arbitrary incident fields, as a result from unequal beam splitting ratios ($t_\mathrm{bs} \neq r_\mathrm{bs}$), the outgoing fields are given as 
\begin{eqnarray}
b_1= r r_\mathrm{bs} e^{i (2 k l_\mathrm{a} +\Theta_{12})}a_\mathrm{in}+t t_\mathrm{bs} e^{i (k(l_\mathrm{a}+l_\mathrm{b}) +\Theta_{13})}a_\mathrm{in},\\
b_2= r t_\mathrm{bs} e^{i (2 k l_\mathrm{b} +\Theta_{13})}a_\mathrm{in}+t r_\mathrm{bs} e^{i (k(l_\mathrm{a}+l_\mathrm{b}) +\Theta_{12})}a_\mathrm{in}.
\end{eqnarray}

\begin{figure}[ht]
\begin{center}
\includegraphics[width=12cm]{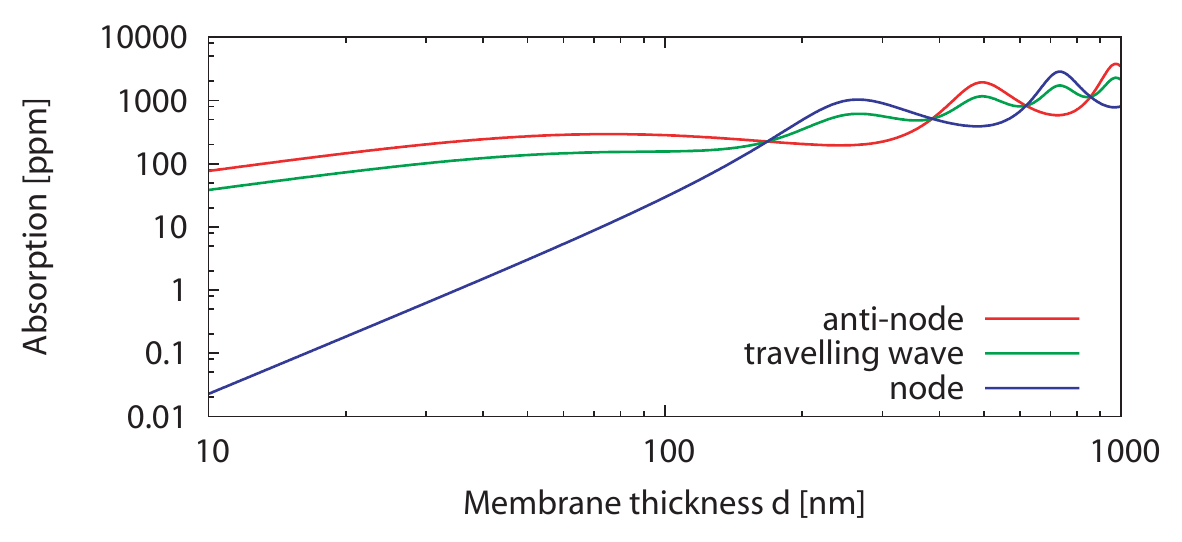}
\caption{Absorption coefficient of SiN-membranes with index of refraction of $n_2=2.2+i1.5\times 10^{-4}$ at a laser wavelength of $\lambda=1064\,$nm \cite{Jayich08} depending on its geometrical thickness $d$. The impact of a placing the membrane in a node and anti-node of a standing wave as well as in travelling wave are compared.} \label{fig:absorption}
\end{center}
\end{figure}

The absorption loss can then be derived by
\begin{eqnarray}
A=1-\left( |b_1|^2+|b_2|^2 \right)/|a_\mathrm{in}|^2\\
= 1-\left[ |r|^2 +|t|^2 + (r^* t +t^* r) 2 r_\mathrm{bs} t_\mathrm{bs} \cos(k(\Delta l)+\Theta_{12}-\Theta_{13}) \right], \label{eq:absorption}
\end{eqnarray}
which in case of $r_\mathrm{bs}=0$, $t_\mathrm{bs}=1$ is the absorption of a travelling wave, which is independent of the mirror position. In Fig.~\ref{fig:absorption} the absorption loss given by Eq.~(\ref{eq:absorption}) is exemplified by a plot drawn for a SiN membrane with index of refraction $n_2=2.2+i1.5\times 10^{-4}$ at a laser wavelength of $\lambda=1064\,$nm \cite{Jayich08} dependent on its geometrical thickness $d$. 
Here, we have chosen a perfect 50/50 beam splitter, which gives the largest difference for a mirror placed at node and anti-node, given by Eq.~(\ref{eq:node}) and Eq.~(\ref{eq:antinode}), respectively. In particular for membranes with thickness  $d\ll \lambda$ optical absorption can be significantly lower if being placed at a node.
 
\section{Experimental results with a SiN-membrane}
We experimentally investigated the impact of high laser power on a translucent mechanical oscillator in a Michelson-Sagnac interferometer. Namely we investigated frequency shifts and a corresponding decrease in the membrane's mechanical quality factor caused by optical absorption. For this purpose we have chosen a commercially available low-stress SiN membrane \cite{norcada}.  

%Their optical absorption has been accurately determined in high finesse cavity experiments and was presented in terms of the imaginary part of index of refraction $Im(n_2)\leq 1.5e-4$\cite{Jayich08}. 

\begin{figure}[ht]
\begin{center}
\includegraphics[width=12cm]{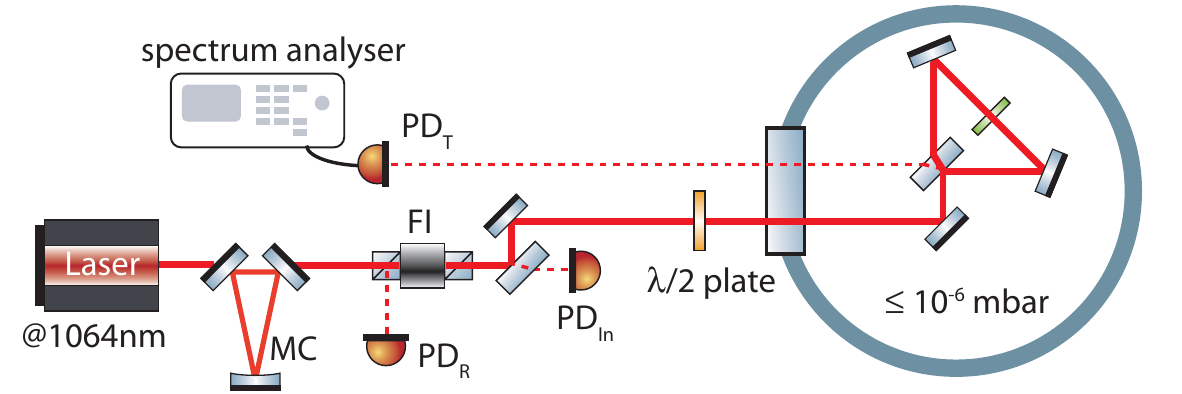}
\caption{Experimental setup including spatial filtering via a triangular modecleaning cavity (MC) and calibrated photodiodes to measure incident (PD$_\mathrm{in}$), transmitted (PD$_\mathrm{T}$), and reflected (PD$_\mathrm{R}$) light power simultaneously. The Michelson-Sagnac interferometer (Michelson-Sagnac ifo) was  operated at a pressure of $\leq 10^{-6}\,$mbar, which was sufficient to exclude viscous damping of the membrane's motion.} \label{fig:experiment}
\end{center}
\end{figure}

The experimental setup is sketched in Fig.~\ref{fig:experiment}. 
The interferometer including a $66\,$nm thin SiN membrane with a measured power transmissivity of $t_\mathrm{m}^2=0.696$ at a laser wavelength of $1064\,$nm was set up in a vacuum environment $\leq 10^{-6}\,$mbar. This enabled us to study the membrane motion unaffected of viscous damping caused by residual gas. The laser preparation, namely spatial filtering via a triangular cavity (MC) \cite{Willke98} as well as a mode matching telescope were set up outside the vacuum chamber. The beam diameter on the membrane with area of ($D^2=1.5 \times 1.5$)mm$^2$ was $\approx 500\,\mu$m. The incident (PD$_\mathrm{In}$), reflected (PD$_\mathrm{R}$), and transmitted light power (PD$_\mathrm{T}$) with respect to the interferometer were measured with calibrated photodiodes. By placing the membrane at the center of the interferometer ($l_\mathrm{a}=l_\mathrm{b}$) we achieved a high interference contrast with losses of $\approx 400\,$ppm (ratio of reflected (PD$_\mathrm{R}$) and transmitted power (PD$_\mathrm{T}$) at a dark fringe).

\begin{figure}[ht]
\begin{center}
\includegraphics[width=12cm]{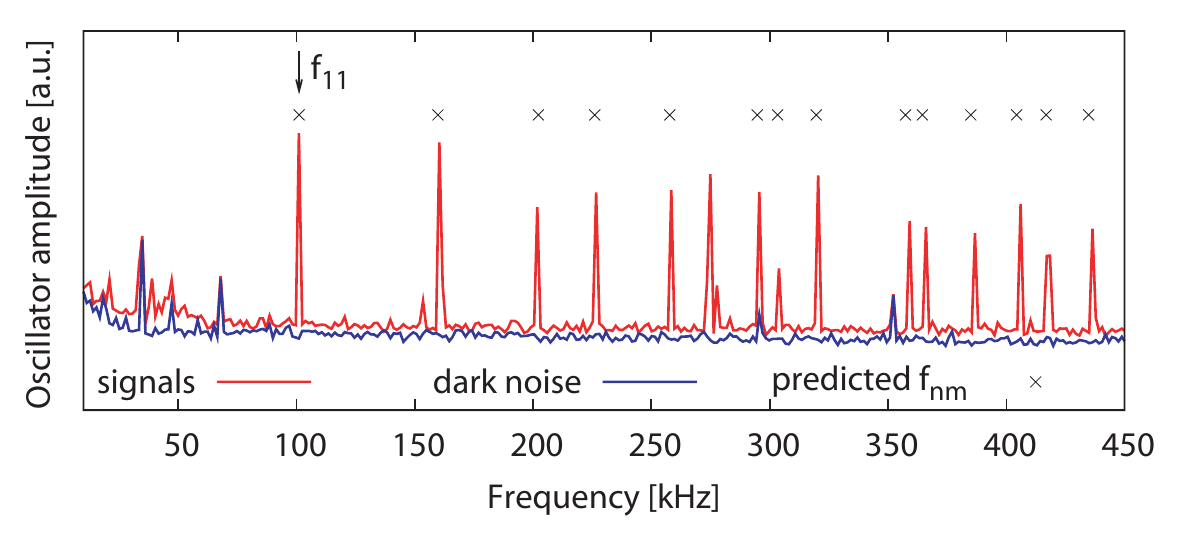}
\caption{Measured spectrum of eigenfrequencies $f_{pq}$ for our SiN-membrane and predicted values with the parameters given in Eq.~(\ref{eq:eigenfrequency}). Note, that the widths of the resonances were dominated by the resolution bandwidth of the spectrum analyzer.} \label{fig:spectrum}
\end{center}
\end{figure}
In a first measurement we confirmed that the eigenfrequencies of a square membrane with side length $D$ are well predicted by 
\begin{eqnarray}
f_{pq}=\frac{1}{2}\sqrt{\frac{\sigma}{\rho}\left( \frac{p^2+q^2}{D^2}  \right)}, \label{eq:eigenfrequency}
\end{eqnarray}
where $\sigma$ is the tensile stress \cite{Wilson11}. Assuming a material density of $\rho=2.7\,$g/cm$^3$ we find  $\sigma \approx 129\,$MPa for the sample under investigation with its fundamental mode at $f_{11}\approx 103\,$kHz. The spectra of measured and predicted eigenfrequencies are in good agreement as shown in Fig.~\ref{fig:spectrum}. According to Eq.~(\ref{eq:eigenfrequency}) one can expect that heating the membrane (neglecting effects from the frame) via optical absorption will result in a decreased eigenfrequency due to expansion and thus lower stress. 

\begin{figure}[ht]
\begin{center}
\includegraphics[width=12cm]{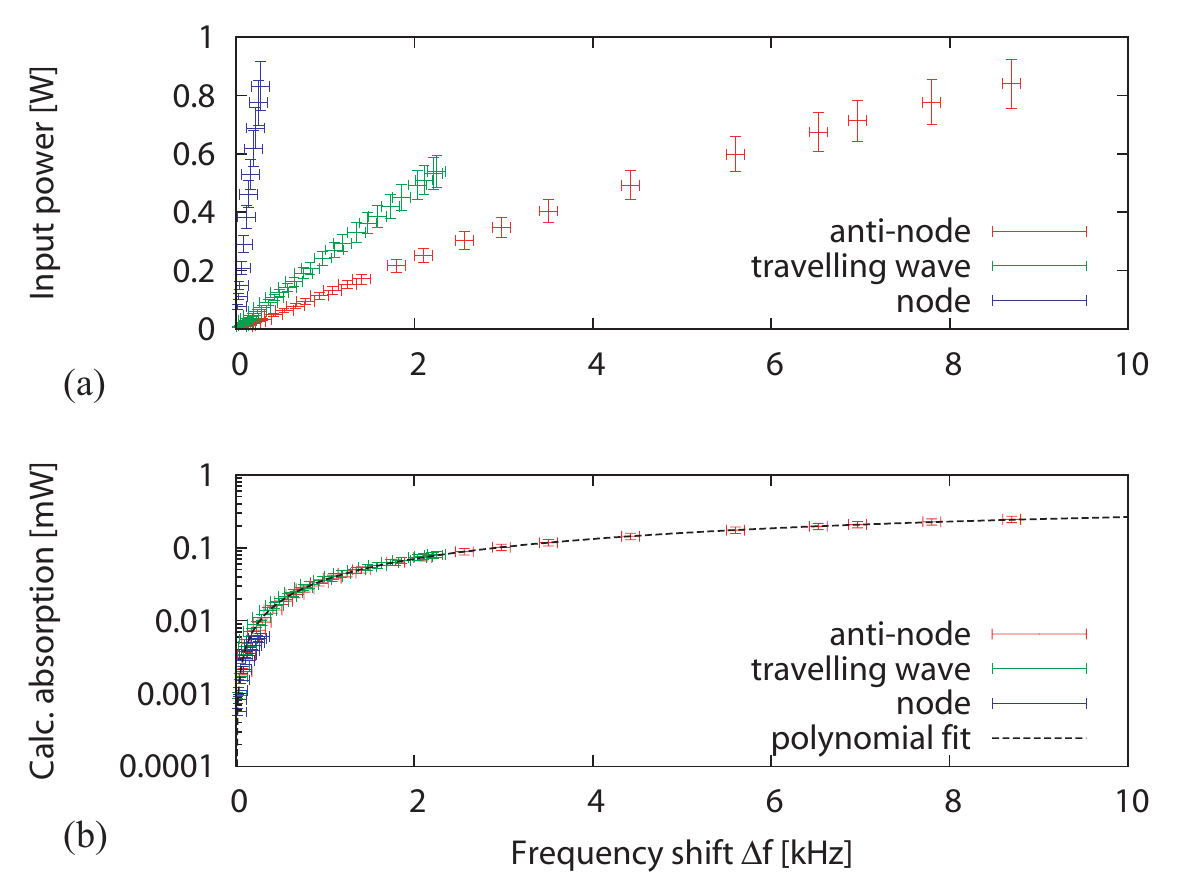}
\caption{(a) Measured shifts of the membrane eigenfrequency $\Delta f$ for different light powers at the optical node and anti-node of a standing wave in the Michelson-Sagnac interferometer as well as for a travelling wave. (b) Corresponding optical absorption calculated via Eq.~(\ref{eq:absorption}), giving $A_{\Delta f}=0.0373\Delta f-0.0011\Delta f^2$ via a polynomial fit.} \label{fig:eigenfrequency}
\end{center}
\end{figure}

\begin{figure}[ht]
\begin{center}
\includegraphics[width=12cm]{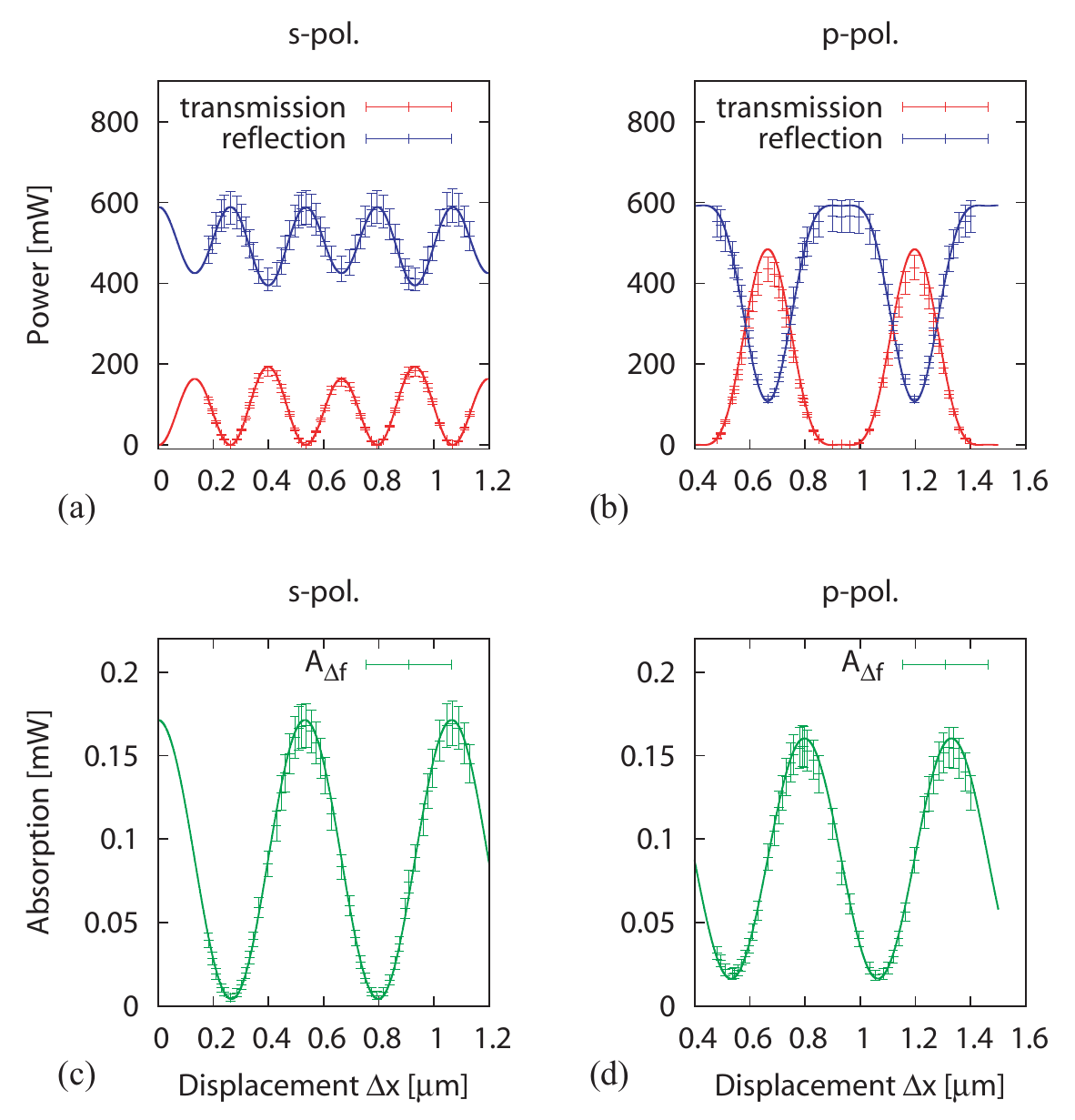}
\caption{(a),(b) Measured and predicted (solid line) output powers of the Michelson-Sagnac interferometer (s- and p-polarized incident light) for a membrane displacement $\Delta x$ and constant incident power of $\approx 600\,$mW. (c),(d) Corresponding optical absorption derived by measured frequency shifts and applying Eq.~(\ref{eq:Adeltaf}). The lowest optical absorption can be achieved for an interferometer with a close to 50/50 beam splitter operated at a particular dark fringe.} \label{fig:expfringes}
\end{center}
\end{figure}

The beam splitter used in the experiment had a splitting ratio of $r_\mathrm{bs}^2/t_\mathrm{bs}^2=0.486/0.514$ for s-polarized light, which allowed us to measure frequency shifts for a membrane positioned at an optical node and anti-node for the interferometer operated close to its dark fringe. The impact of a travelling wave was measured by misaligning one of the interferometer's steering mirrors. The light reflected from the membrane was then brought to interference with an auxilary beam outside the vacuum chamber, forming an Michelson interferometer (not shown in Fig.~\ref{fig:experiment}), in order to measure the membrane's eigenfrequency. The measured decrease in frequency $\Delta f$ for varying input power is shown in Fig.~\ref{fig:eigenfrequency}(a). 
Based on Eq.~(\ref{eq:absorption}) we calculated the corresponding optical absorption (see Fig.~\ref{fig:eigenfrequency}(b)). As a result we determined the relation between frequency shift and optical absorption $A_{\Delta f}$ via a polynomial fit to be  
\begin{eqnarray}
A_{\Delta f}=0.0373\Delta f-0.0011\Delta f^2 \label{eq:Adeltaf}
\end{eqnarray}  
for the given membrane and beam size. Here, we have assumed that optical properties do not change for larger absorption. In order to compare the theoretical predictions for the position dependent absorption in the Michelson-Sagnac interferometer, namely the cosine dependence of Eq.~(\ref{eq:absorption}), we have measured the interferometer's output power (PD$_\mathrm{R}$, PD$_\mathrm{T}$) for different membrane displacements $\Delta x$ (see Fig.~\ref{fig:expfringes}(a)(b)) and corresponding frequency shifts.  Based on Eq.~(\ref{eq:Adeltaf}) we related the measured position dependent frequency shifts to optical absorption, which is shown in Fig.~\ref{fig:expfringes}(c)(d). The measurements were carried out for s- and p-polarized light to test the theoretical predictions for different splitting ratios of the central beam splitter $r_\mathrm{bs}^2/t_\mathrm{bs}^2$, which were measured to be $0.486/0.514$ and $0.241/0.759$, respectively.  

\begin{figure}[ht]
\begin{center}
\includegraphics[width=12cm]{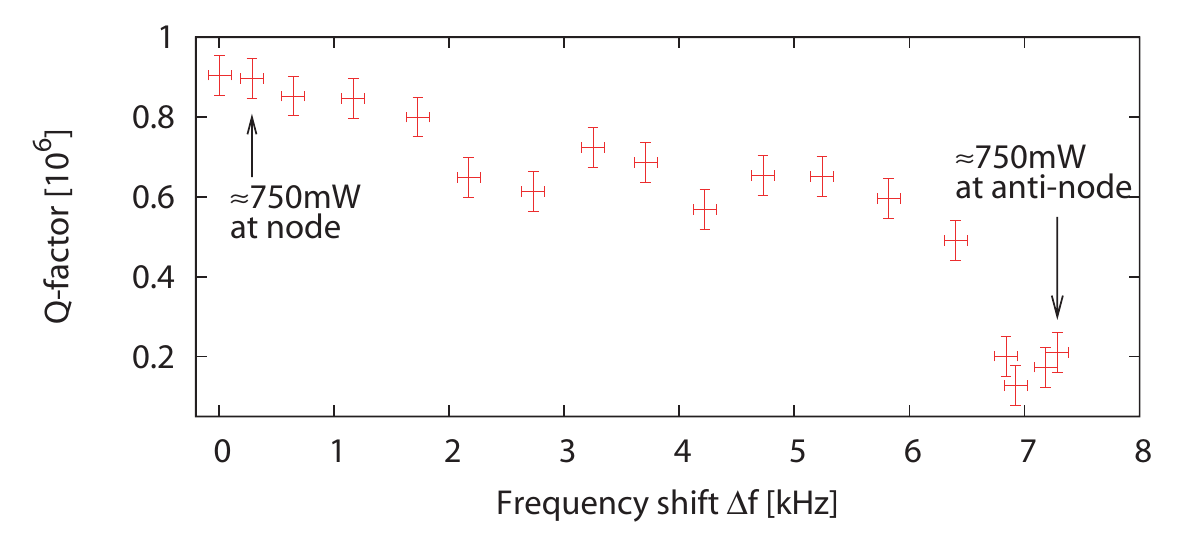}
\caption{Measured mechanical quality factors for a range of negative frequency shifts $\Delta f$ caused by optical absorption with increased laser power. Quality factors were determined from ring-down time measurements.} \label{fig:Qfactor}
\end{center}
\end{figure}

The experimental results demonstrate that the lowest optical absorption can be achieved in a Michelson-Sagnac interferometer operated at its dark fringe for a close to 50/50 splitting ratio of the central beam splitter. The absorption for p-polarized light was less pronounced compared to s-polarized light due to the existence of a partial standing wave. We further measured the impact of optical absorption on the mechanical quality factor. Ring-down time measurements of the membrane's amplitude after an excitation were taken for various frequency shifts. The results in Fig.~\ref{fig:Qfactor} show a significant decrease in the membrane's mechanical quality factor with increasing absorption by about an order of magnitude. Each point shown was averaged over five measurements. 
Although the presented decrease in mechanical quality factor is specific for the given setup it emphasizes that optical absorption needs to be minimized, which is possible with the interferometer topology investigated here.

\section{Conclusion}
We have realized a Michelson-Sagnac interferometer containing a SiN membrane as a high-quality mechanical oscillator. We have measured the frequency and the mechanical quality of the membrane's fundamental mode of motion and found a pronounced influence of the membrane's microscopic position within the interferometer. By placing the membrane into the node of the standing optical wave of the interferometer, lowest values for absorption and quality factor degradation have been observed. Our optical models have been found to be in excellent agreement with our experimental data. Since the coupling of a light field to the motion of a mechanical oscillator via radiation pressure does not depend on the light power \textit{at} the position of the oscillator but on the power \textit{reflected} from the oscillator, our setup provides a promising approach for quantum opto-mechanical experiments with oscillators showing substantial light transmission and also optical absorption.

\section*{Acknowledgements}
This work is supported by the Excellence Cluster QUEST and the IMPRS on Gravitational Wave Astronomy. S. D. is supported by the Alexander von Humboldt Foundation.


\section*{References}
\begin{thebibliography}{25}

\bibitem{Braginsky78} Braginsky V B, Vorontsov Yu I and Khalili F Ya 1978 Optimal quantum measurements in gravitational-wave detectors {\it Pisma v Zh. Eksp. Teor. Phys} {\bf 27} 296--301

\bibitem{Thorne78} Thorne K S, Drever R W P, Caves C M, Zimmermann M and Sandberg V D 1978 Quantum Nondemolition Measurements of Harmonic Oscillators {\it Phys. Rev. Lett.} {\bf 40} 667--671

\bibitem{Braginsky} Braginsky V B and Khalili F Ya {\it Quantum measurement} 1992 (Cambridge: Cambridge University Press)

\bibitem{Chen11} Chen Y, Danilishin S L, Khalili F Ya, M\"uller-Ebhardt H 2011 QND measurements for future gravitational-wave detectors {\it General Relativity and Gravitation} {\bf 43} 671--694

\bibitem{AdvLIGO} Harry G M (for the LIGO Scientific Collaboration) 2010 Advanced LIGO: the next generation of gravitational wave detectors {\it Class. Quantum Grav.} {\bf 27} 084006

\bibitem{AdvVirgo} Acernese F et al. Advanced Virgo baseline design VIR-027A-09 (https://tds.ego-gw.it/ql/?c=6589)

\bibitem{GeoHF} Willke B et al. 2006 The GEO-HF project {\it Class. Quantum Grav.} {\bf 23} S207--S214

\bibitem{LCGT} Kuroda K (on behalf of the LCGT Collaboration) 2010 Status of LCGT {\it Class. Quantum Grav.} {\bf 27} 084004

\bibitem{Caves80} Caves C M 1980 Quantum-Mechanical Radiation-Pressure Fluctuations in an Interferometer {\it Phys. Rev. Lett.} {\bf 45} 75--79

\bibitem{Kippenberg08} Kippenberg T J and Vahala K J 2008 Cavity Optomechanics: Back-Action at the Mesoscale {\it Science} {\bf 321} 172--176

\bibitem{Levin98} Levin Yu 1998 Internal thermal noise in the LIGO test masses: A direct approach {\it Phys. Rev.} D {\bf 57} 659--663

\bibitem{Harry02} Harry G M, Gretarsson A M, Saulson P R, Kittelberger S E, Penn S D, Startin W J, Rowan S, Fejer M M, Crooks D R M, Cagnoli G, Hough J, and Nakagawa N 2002 Thermal noise in interferometric gravitational wave detectors due to dielectric optical coatings {\it Class. Quantum Grav.} {\bf 19} 897--917

\bibitem{Thompson08} Thompson J D, Zwickl B M, Jayich A M, Marquardt F, Girvin S M and Harris J G E 2008 Strong dispersive coupling of a high-finesse cavity to a micromechanical membrane {\it Nature} {\bf 452} 72--75

\bibitem{Sankey10} Sankey J C, Yang C, Zwickl B M, Jayich A M and Harris J G E 2010 Strong and tunable nonlinear optomechanical coupling in a low-loss system {\it Nature Physics} {\bf 6} 707--712

\bibitem{Jayich08} Jayich A M, Sankey J C, Zwickl B M, Yang C, Thompson J D, Girvin S M, Clerk A A, Marquardt F and Harris J G E 2008 Dispersive optomechanics: a membrane inside a cavity {\it New Journal of Physics} {\bf 10} 095008

\bibitem{Zwickl08} Zwickl B M, Shanks W E, Jayich A M, Yang C, Bleszynski Jayich A C, Thompson J D and Harris J G E 2008 High quality mechanical and optical properties of commercial silicon nitride membranes {\it Applied Physics Letters} {\bf 92} 103125

\bibitem{Meers88} Meers B J 1988 Recycling in laser-interferometric gravitational-wave detectors {\it Phys. Rev.} D {\bf 38} 2317--2326

\bibitem{Strain91} Strain K A and Meers B J 1991 Experimental Demonstration of Dual Recycling for Interferometric Gravitational-Wave Detectors {\it Phys. Rev. Lett.} {\bf 66} 1391--1394

\bibitem{Yamamoto10} Yamamoto K, Friedrich D, Westphal T, Go\ss ler S, Danzmann K, Somiya K, Danilishin S L and Schnabel R 2010 Quantum noise of a Michelson-Sagnac interferometer with a translucent mechanical oscillator {\it Phys. Rev.} A {\bf 81} 033849

\bibitem{Zeilinger81} Zeilinger A 1981 General properties of lossless beam splitters in interferometry {\it Am. J. Phys.} {\bf 49} 882--883

\bibitem{Born} M. Born and E. Wolf 'Principles of Optics' 1970 (Oxford: Pergamon press)

\bibitem{norcada} www.norcada.com

\bibitem{Willke98} Willke B, Uehara N, Gustafson E K, Byer R L, King P J, Seel S U and Savage Jr R L 1998 Spatial and temporal filtering of a 10-W Nd:YAG laser with a Fabry-Perot ring-cavity premode cleaner {\it Opt. Lett.} {\bf 23} 1704--1706

\bibitem{Wilson11} Wilson-Rae I, Barton R A, Verbridge S S, Southworth D R, Ilic B, Craighead H G and Parpia J M 2011 High-Q Nanomechanics via Destructive Interference of Elastic Waves {\it Phys. Rev. Lett.} {\bf 106} 047205

\end{thebibliography}
\end{document}